\documentclass[10pt,a4paper,onecolumn]{article}
\usepackage{marginnote}
\usepackage{graphicx}
\usepackage{xcolor}
\usepackage{authblk,etoolbox}
\usepackage{titlesec}
\usepackage{calc}
\usepackage{tikz}
\usepackage{hyperref}
\hypersetup{colorlinks,breaklinks,
            urlcolor=[rgb]{0.0, 0.5, 1.0},
            linkcolor=[rgb]{0.0, 0.5, 1.0}}
\usepackage{caption}
\usepackage{tcolorbox}
\usepackage{amssymb,amsmath}
\usepackage{ifxetex,ifluatex}
\usepackage{seqsplit}
\usepackage{fixltx2e} % provides \textsubscript
\usepackage[
  backend=biber,
]{biblatex}
\usepackage[top=3.5cm, bottom=3cm, right=1.5cm, left=1.0cm,
            headheight=2.2cm, reversemp, includemp, marginparwidth=4.5cm]{geometry}

\titleformat{\section}
  {\normalfont\sffamily\Large\bfseries}
  {}{0pt}{}
\titleformat{\subsection}
  {\normalfont\sffamily\large\bfseries}
  {}{0pt}{}
\titleformat{\subsubsection}
  {\normalfont\sffamily\bfseries}
  {}{0pt}{}
\titleformat*{\paragraph}
  {\sffamily\normalsize}

\usepackage{fancyhdr}
\makeatletter
\let\ps@plain\ps@fancy
\fancyheadoffset[L]{4.5cm}
\fancyfootoffset[L]{4.5cm}

\definecolor{linky}{rgb}{0.0, 0.5, 1.0}

\newtcolorbox{repobox}
   {colback=red, colframe=red!75!black,
     boxrule=0.5pt, arc=2pt, left=6pt, right=6pt, top=3pt, bottom=3pt}

\newcommand{\ExternalLink}{%
   \tikz[x=1.2ex, y=1.2ex, baseline=-0.05ex]{%
       \begin{scope}[x=1ex, y=1ex]
           \clip (-0.1,-0.1)
               --++ (-0, 1.2)
               --++ (0.6, 0)
               --++ (0, -0.6)
               --++ (0.6, 0)
               --++ (0, -1);
           \path[draw,
               line width = 0.5,
               rounded corners=0.5]
               (0,0) rectangle (1,1);
       \end{scope}
       \path[draw, line width = 0.5] (0.5, 0.5)
           -- (1, 1);
       \path[draw, line width = 0.5] (0.6, 1)
           -- (1, 1) -- (1, 0.6);
       }
   }

\patchcmd{\@maketitle}{center}{flushleft}{}{}
\patchcmd{\@maketitle}{center}{flushleft}{}{}
\patchcmd{\@maketitle}{\LARGE}{\LARGE\sffamily}{}{}
\def\maketitle{{%
  
  \AB@maketitle}}
\makeatletter
\renewcommand\AB@affilsepx{ \protect\Affilfont}
\renewcommand\AB@affilnote[1]{{\bfseries #1}\hspace{3pt}}
\makeatother

\renewcommand\Affilfont{\sffamily\small\mdseries}
\ifnum 0\ifxetex 1\fi\ifluatex 1\fi=0 % if pdftex
  \usepackage[T1]{fontenc}
  \usepackage[utf8]{inputenc}

\else % if luatex or xelatex
  \ifxetex
    \usepackage{mathspec}
  \else
    \usepackage{fontspec}
  \fi
  \defaultfontfeatures{Ligatures=TeX,Scale=MatchLowercase}

\fi
\IfFileExists{upquote.sty}{\usepackage{upquote}}{}
\IfFileExists{microtype.sty}{%
\usepackage{microtype}
\UseMicrotypeSet[protrusion]{basicmath} % disable protrusion for tt fonts
}{}

\usepackage{hyperref}
\hypersetup{unicode=true,
            pdftitle={hankel: A Python library for performing simple and accurate Hankel transformations},
            pdfborder={0 0 0},
            breaklinks=true}
\usepackage{graphicx,grffile}
\makeatletter
\def\maxwidth{\ifdim\Gin@nat@width>\linewidth\linewidth\else\Gin@nat@width\fi}
\def\maxheight{\ifdim\Gin@nat@height>\textheight\textheight\else\Gin@nat@height\fi}
\makeatother
\setkeys{Gin}{width=\maxwidth,height=\maxheight,keepaspectratio}
\IfFileExists{parskip.sty}{%
\usepackage{parskip}
}{% else
\setlength{\parindent}{0pt}
\setlength{\parskip}{6pt plus 2pt minus 1pt}
}
\ifx\paragraph\undefined\else
\let\oldparagraph\paragraph
\renewcommand{\paragraph}[1]{\oldparagraph{#1}\mbox{}}
\fi
\ifx\subparagraph\undefined\else
\let\oldsubparagraph\subparagraph
\renewcommand{\subparagraph}[1]{\oldsubparagraph{#1}\mbox{}}
\fi

\title{hankel: A Python library for performing simple and accurate Hankel
transformations}

        \author[1, 2, 3]{Steven G. Murray}
          \author[4]{Francis J. Poulin}
    
      \affil[1]{International Centre for Radio Astronomy Research (ICRAR), Curtin
University, Bentley, WA 6102, Australia}
      \affil[2]{ARC Centre of Excellence for All-Sky Astrophysics in 3 Dimensions (ASTRO
3D)}
      \affil[3]{School of Earth and Space Exploration, Arizona State University, Tempe,
AZ, 85281, USA}
      \affil[4]{Department of Applied Mathematics, University of Waterloo}
  \date{\vspace{-5ex}}

\begin{document}
\maketitle

\marginpar{
  %\hrule
  \sffamily\small

  {\bfseries DOI:} \href{https://doi.org/10.21105/joss.01397}{\color{linky}{10.21105/joss.01397}}

  \vspace{2mm}

  {\bfseries Software}
  \begin{itemize}
    \setlength\itemsep{0em}
    \item \href{https://github.com/openjournals/joss-reviews/issues/37}{\color{linky}{Review}} \ExternalLink
    \item \href{https://github.com/steven-murray/hankel}{\color{linky}{Repository}} \ExternalLink
    \item \href{http://dx.doi.org/10.21105/joss.01397}{\color{linky}{Archive}} \ExternalLink
  \end{itemize}

  \vspace{2mm}

  {\bfseries Submitted:} 02 April 2019\\
  {\bfseries Published:} 31 May 2019

  \vspace{2mm}
  {\bfseries License}\\
  Authors of papers retain copyright and release the work under a Creative Commons Attribution 4.0 International License (\href{http://creativecommons.org/licenses/by/4.0/}{\color{linky}{CC-BY}}).
}

\hypertarget{summary}{%
\section{Summary}\label{summary}}

The Hankel transform is a one-dimensional functional transform involving
a Bessel-function kernel. Moreover, it is the radial solution to an
angularly symmetric Fourier transform of any dimension, rendering it
very useful in several fields of application. The NASA Astronomical Data
Service yields over 700 refereed articles including the term ``hankel
transform'', in fields as diverse as astronomy, geophysics, fluid
mechanics, electrodynamics, thermodynamics and acoustics.

As an example, in cosmology, the density field of the Universe is
expected to be isotropic. One of the primary means of describing this
field is via its Fourier power spectrum, or equivalently its spatial
autocorrelation function. Due to the isotropy of the field, these can be
related by an angularly symmetric Fourier transform, which is more
simply expressed as a Hankel transform (Szapudi et al. 2005).

Another example that arises in both geophysical and astrophysical
contexts is in regards to vortices. The radially-symmetric vortical
solution to Laplace's equation in two, three or even higher dimensions
can be performed quickly and accurately via the Hankel transform (Carton
2001).

Conceptually, computation of such problems using the Hankel transform,
in contrast to the Fourier transform, has the advantage of reducing the
problem's dimensionality to unity, regardless of the original
dimensionality. Analytically, this \emph{may} be a useful tool in
solving the transform. Numerically, it naively promises to enhance
efficiency.

Despite these advantages, the Hankel transform introduces some numerical
challenges. Most importantly, the Hankel transform is a \emph{highly
oscillatory} integral, especially for large values of the transformation
variable, \emph{k} (henceforth we will use \emph{r} to denote the
magnitude of the real-space co-ordinate). Highly oscillatory integrals
are a topic of much interest in applied mathematics, and there does not
exist a general optimal solution to numerically evaluate them in general
(Huybrechs and Olver 2009). Nevertheless, Ogata (2005) determined that a
double-exponential variable transformation based on the zeros of the
Bessel function (Ooura and Mori 1999) has the property that the
numerical integral converges with many fewer divisions compared to
naively computing the transform integral. This procedure is able to
efficiently and accurately evaluate the Hankel integral (and hence the
Hankel transform) in many cases.

The purpose of \texttt{hankel} is to provide a dead-simple intuitive
pure-Python interface for performing Hankel integrals and transforms,
written with both Python 2 and 3 compatibility. It enables the accurate
determination of the transform of a callable function at any arbitrary
value of the transform variable, \emph{k}, and utilises the proven
method of Ogata (2005) to do this efficiently. In addition, it
recognizes that the most common application of the Hankel transform is
in the context of the radial Fourier transform, and it provides an
additional interface dedicated to making the connection between these
transforms more transparent.

The chief performance-critical components of \texttt{hankel} are the
evaluation of the zeros of the Bessel function, and the sum of terms
required for integration. The former is made efficient by a tiered
approach -- using the efficient \texttt{scipy} for integer-order Bessel
functions, directly returning regular arrays for order-1/2, and using
the very accurate \texttt{mpmath} for all other orders. The latter is
made efficient by utilising \texttt{numpy}, bringing it close to C-level
performance.

The \texttt{hankel} package is thoroughly tested to ensure accuracy of
transforms, by comparing to known analytic solutions. These tests,
supported by continuous integration, are also useful for the user who
wishes to explore the numerical limitations of the method. Aside from
functions which are theoretically divergent, the method can struggle to
transform several classes of functions, including those with very sharp
features, especially at small \emph{r}. The method itself has two free
parameters, \emph{h} and \emph{N}, which respectively determine the
resolution and upper limit of the integration grid. These can be
modified to accurately transform any function that theoretically
converges. How to choose these values, and the estimated error of the
transform under a given choice, are discussed in the \texttt{hankel}'s
extensive online documentation (and the reader is referred to Ogata
(2005) for more details). Based on the arguments in the documentation,
\texttt{hankel} provides an automatic, guided-adaptive algorithm for
determination of \emph{h} and \emph{N}.

A particularly important limitation of \texttt{hankel}, as currently
implemented, is that it does \emph{not} implement the \emph{discrete
Hankel transform}. That is, it provides no direct means of transforming
an array of regular-spaced function values into ``radial Fourier space''
at regular-spaced \emph{k} values. It is focused solely on transforming
\emph{callable} functions, so that it can evaluate that function at the
non-regular locations required by the double-exponential transform of
Ogata (2005). Extensions to discrete Hankel transforms (and even
\emph{fast} Hankel transforms) are envisioned for v2.0 of
\texttt{hankel}.

\hypertarget{acknowledgements}{%
\section{Acknowledgements}\label{acknowledgements}}

The authors acknowledge helpful contributions from Sebastian Mueller
during the construction of this code. Parts of this research were
supported by the Australian Research Council Centre of Excellence for
All Sky Astrophysics in 3 Dimensions (ASTRO 3D), through project number
CE170100013. FJP would like to thank NSERC for research funding during
the time of this research.

\hypertarget{references}{%
\section*{References}\label{references}}
\addcontentsline{toc}{section}{References}

\hypertarget{refs}{}
\leavevmode\hypertarget{ref-carton2001hydrodynamical}{}%
Carton, Xavier. 2001. ``Hydrodynamical Modeling of Oceanic Vortices.''
\emph{Surveys in Geophysics} 22 (3): 179--263.

\leavevmode\hypertarget{ref-huybrechs_olver_2009}{}%
Huybrechs, D., and S. Olver. 2009. ``Highly Oscillatory Quadrature.'' In
\emph{Highly Oscillatory Problems}, 25--50. London Mathematical Society
Lecture Note Series. Cambridge University Press.
\url{https://doi.org/10.1017/CBO9781139107136.003}.

\leavevmode\hypertarget{ref-Ogata2005}{}%
Ogata, Hidenori. 2005. ``A Numerical Integration Formula Based on the
Bessel Functions.'' \emph{Publ RIMS Kyoto Univ} 41: 949--70.
\url{https://doi.org/10.2977/prims/1145474602}.

\leavevmode\hypertarget{ref-Ooura1999}{}%
Ooura, Takuya, and Masatake Mori. 1999. ``A Robust Double Exponential
Formula for Fourier-Type Integrals.'' \emph{Journal of Computational and
Applied Mathematics} 112 (1): 229--41.
\url{https://doi.org/10.1016/S0377-0427(99)00223-X}.

\leavevmode\hypertarget{ref-Szapudi2005}{}%
Szapudi, István, Jun Pan, Simon Prunet, and Tamás Budavári. 2005. ``Fast
Edge Corrected Measurement of the Two-Point Correlation Function and the
Power Spectrum.'' \emph{Arxiv E-Prints}, May, 4--4.
\url{https://doi.org/10.1086/496971}.

\end{document}